\documentstyle[preprint,epsfig,eqsecnum,aps]{revtex}

\begin{document}

\title{The Single State Dominance Hypothesis and the Two-Neutrino Double Beta 
       Decay of $^{100}Mo$}

\author{F. \v Simkovic$^1$, 
P. Domin$^1$, and S.V. Semenov$^{2}$}
\address{
1. Department of Nuclear Physics, Comenius University, Bratislava, Slovakia\\
2. Russian Research Center "Kurchatov Institute", Moscow, Russia}

\date{\today}

\maketitle


\date{\today}
\maketitle


\begin{abstract}
The hypothesis of the single state dominance (SSD) in the calculation of the
two-neutrino double beta decay ($2\nu\beta\beta$-decay)	
of $^{100}Mo$ is tested by exact
consideration of the energy denominators of the perturbation theory.
Both transitions to the ground state as well as to the $0^+$ and
$2^+$ excited states of the final nucleus $^{100}Ru$ are
considered. We demonstrate, that by experimental investigation 
of the single electron energy distribution and the angular correlation
of the outgoing electrons,  the SSD hypothesis 
can be  confirmed or ruled out  by a precise
$2\nu\beta\beta$-decay measurement (e.g. by NEMO III collaboration). 

\vskip0.1cm

\noindent 
{\it PACS number(s):} 21.60.Jz; 23.40.Bw; 23.40.Hc
\end{abstract}

\section{Introduction}
\label{sec:level1}

The two-neutrino double beta decay ($2\nu\beta\beta$-decay) 
\cite{HAX84,DOI85,tom91}, which is allowed by the Standard model, has been 
 observed in direct counter experiments  
for a couple of isotopes during the last 15 years
(see e.g. the recent review articles \cite{fae98,suh98}). As
the decay rate of this process is free of unknown parameters from
the particle physics side this very rare process with a typical half-life
above $10^{18}$ years can be used to test the nuclear structure. 

The $2\nu\beta\beta$-decay is a second order process in 
perturbation theory. Thus the calculation of the $2\nu\beta\beta$-decay 
matrix element is a complex task mainly due to the fact that  it involves 
a summation over a full set of virtual intermediate  nuclear states
of the double--odd nucleus. Essentially there  exist two different 
approaches to evaluate the $2\nu\beta\beta$-decay rate including an explicit 
summation over these states \cite{fae98,suh98}: 
The shell model approach has been found successful in describing 
satisfactory the lowest excited states, however, it  can not reliably 
describe the states in the giant Gamow-Teller resonance region
for open shell medium heavy nuclei. The proton-neutron
QRPA (pn-QRPA) and its extensions avoid this drawback but on other hand 
their  predictions are very dependent upon model assumptions.

The crucial problem of the theoretical $2\nu\beta\beta$-decay studies is 
the question whether the contribution of the higher-lying states
to the $2\nu\beta\beta$-decay amplitude, which
is apparently disfavored by the large energy denominator, 
plays an important role. In Ref. \cite{abad84} it was suggested 
that $2\nu\beta\beta$-decay transitions,
where the first $1^+_1$ state of the intermediate nucleus 
($A,Z\pm 1$) is the ground state, are  governed only by the 
following two beta transitions:
i) The first one connecting the ground state 
of the initial nucleus ($A,Z$) with  $1^+_1$ intermediate state of
$(A,Z\pm 1)$ nucleus.
ii) The second one proceeding from the $1^+_1$ state to the final ground state 
($A,Z\pm 2$). 
This assumption is known as the {\em single state dominance(SSD) hypothesis}.
We note that the dominance of the ground state of intermediate nucleus
in particular case of $2\nu\beta\beta$-decay of $^{100}Mo$ was 
pointed out by A. Griffiths and P. Vogel, who analysed this decay in
details \cite{vogel}.

The SSD hypothesis has been studied both experimentally 
\cite{abad84,gar93,aki97,bha98} and theoretically \cite{vogel,eji96,civ98}.
The required beta transition amplitudes to the $1^+_1$ state
have been deduced from the measured $log ft$ values, i.e., in the model
independent way, or have been calculated e.g. 
within the pn-QRPA \cite{vogel,civ98}.   
The obtained results indicate that the SSD hypothesis can be realized
in the case of several $2\nu\beta\beta$-decay emitters through a true
dominance of the $1^+_1$ state or by cancelations among the higher lying
$1^+$ state of the intermediate nucleus. Till now the study 
has been  concentrated mostly on the determination of the 
$2\nu\beta\beta$-decay half-life for the transition to the ground state. 
Recently the $2\nu\beta\beta$-decay 
transitions to excited $0^+$ states of the final nucleus has gained
much attention \cite{civ98}. It is worthwhile to notice 
that there is now a first positive evidence for such nuclear 
$2\nu\beta\beta$-decay transition \cite{barb95}.

In  previous SSD hypothesis studies, calculations have been
performed with approximated energy denominators of the perturbation
theory by ignoring  their dependence on lepton energies 
\cite{civ98}. However, this approximation can lead to 
significant overestimation
of the $2\nu\beta\beta$-decay half-life as it was shown in Ref. \cite{sem00}.
Therefore, it is necessary to reconsider the SSD predictions  without
the above approximation. It is also supposed that exact calculations of the 
SSD hypothesis  
with unfactorized nuclear part and integration over 
the phase space of the outgoing
leptons can strongly influence the behavior of some of the 
differential decay rates. Previously, they have not been 
analyzed in the framework of  the SSD hypothesis.
However, they are of current interest due to the prepared NEMO III experiment,
which will allow to perform a precise
measurement of the energy and angular distributions of the outgoing 
electrons \cite{bar99}. 

In this paper we perform exact calculations of the
SSD hypothesis  of $2\nu\beta\beta$-decay of
$^{100}Mo$ for the transitions to the $0^+$ ground state as well
as to excited $0^+$ and $2^+$ states of the final nucleus $^{100}Ru$.
In addition,  we will discuss a possible signal
 in favor of the SSD hypothesis from the 
differential decay rates.

\section{Theory}
\label{sec:level2}

The inverse half-life of the $2\nu\beta\beta$-decay transition to the 
$0^+$ and $2^+$ states of the final nucleus is usually presented 
in the following form \cite{HAX84,DOI85,tom91}:
\begin{equation}
[T^{2\nu}_{1/2} (0^+\rightarrow J^+)]^{-1} 
= G^{2\nu}(J^+) 
|M^{}_{GT}(J^+)|^2,
\label{eq:1}
\end{equation}
where $G^{2\nu}(J^+)$ is the kinematical factor.
The nuclear matrix element $M^{2\nu}_{GT}(J^+)$ can be written as sum
of two matrix elements $M^{SS}_{GT}(J^+)$ and $M^{HS}_{GT}(J^+)$ including
the transitions through the lowest and higher lying states of the 
intermediate nucleus, respectively. We have
\begin{equation}
M_{GT}(J^+)= M^{SS}_{GT}(J^+) + 
M^{HS}_{GT}(J^+),\\
\label{eq:2}
\end{equation}
where
\begin{eqnarray}
M^{SS}_{GT}(J^+) &=& \frac{1}{\sqrt{s}}
 \frac{M^f_1(J^+)M^i_1(0^+)}
{[E_1 - E_i + \Delta ]^s},
\nonumber \\ 
 M^{HS}_{GT}(J^+) &=& \frac{1}{\sqrt{s}}
 \sum_{n=2} \frac{M^f_n(J^+)M^i_n(0^+)}
{[E_n - E_i + \Delta ]^s}
\label{eq:3}
\end{eqnarray}
with
\begin{eqnarray}
M^f_n(J^+) &=& <J^+_f\parallel 
\sum_{{m}}\tau^{{+}}_{{m}}{\sigma}_{{m}} \parallel 1^+_n>,
\nonumber \\
M^i_n(0^+) &=& <1^+_n\parallel 
\sum_{{m}}\tau^{{+}}_{{m}}{\sigma}_{{m}} \parallel 0^+_i>.
\label{eq:4}
\end{eqnarray}
Here, s=1 for J=0 and s=3 for J=2. $|0^+_i>$, $|0^+_f>$ and
$|1^+_n>$ are respectively the wave functions of the initial, final
and intermediate nuclei with corresponding energies $E_i$, $E_f$ and
$E_n$.  $\Delta$ denotes the average energy $\Delta = (E_i - E_f)/2$. 

The SSD hypothesis assumes that the nuclear matrix
element $ M^{HS}_{GT}(J^+)$ is negligible in comparison with 
$ M^{SS}_{GT}(J^+)$, which can be determined in phenomenological
(with help of $log ft$ values) or nuclear model dependent
way.  Knowing the value of $ M^{SS}_{GT}(J^+)$ one can predict the 
$2\nu\beta\beta$-decay half-life with help of Eq. (\ref{eq:1}) 
and compare it with the
measured one. Henceforth we shall denote this approach as SSD1.

The SSD1 $2\nu\beta\beta$-decay half-life is derived in the approximation
in which the sum of two lepton energies in the denominator of
the $2\nu\beta\beta$-decay nuclear matrix element is replaced with
their average value $\Delta$:
\begin{equation}
D(\varepsilon_i,\omega_j)\equiv {E_1 - E_i + \varepsilon_i +\nu_j }, 
\approx {E_1 - E_i + \Delta }, 
\label{eq:5}
\end{equation}
$(i,j ~=~1,2)$.
Here, $\varepsilon_i =\sqrt{k^2_i+m_e^2}$ 
($m_e$ is the mass of electron) and $\nu_j$ are energies of electrons
and antineutrinos, respectively. The main purpose of this approximation
is to factorize the lepton and  nuclear parts in the calculation
of $[T^{2\nu}_{1/2} (0^+\rightarrow J^+)$. However, it is not necessary
to do it within the SSD hypothesis. We note that in the particular case of
$2\nu\beta\beta$-decay of $^{100}Mo$ the value $E_1-E_i$  is negative
(-0.343 MeV) and that there is large difference between the  
minimal (0.168 MeV)
and maximal (3.202 MeV) values of $D(\varepsilon_i,\omega_k)$. It indicates 
that one has to go beyond the above approximation.
The SSD hypothesis approach with exact consideration of the energy
denominators will be denoted hereafter  SSD2. 

From the theoretical point of view one can  discuss also an  alternative
assumption, which is the dominance of the contribution from higher 
order states of the intermediate
nucleus to $2\nu\beta\beta$-decay rate. We shall denote it as {\em higher order
state dominance  (HSD) hypothesis}. We note that 
within this assumption one can factorize 
safely the nuclear part and the 
integration over the phase space. Thus it is expected that 
the behavior of the HSD hypothesis differential
decay rates will differ considerably from those obtained within SSD2, if 
the value of the expression $E_1-E_i+m_e$ is rather small. 
For such  comparison of the SSD2 and
HSD approaches we shall assume
\begin{equation}
M^{HS}_{GT} \approx M^{exp}_{GT} =  
[T^{2\nu-exp}_{1/2} (0^+\rightarrow J^+) ~G^{2\nu}(J^+) ]^{-1/2}. 
\label{eq:6}
\end{equation}
Here, $T^{2\nu-exp}_{1/2} (0^+\rightarrow J^+)$ 
is the measured $2\nu\beta\beta$-decay half-life.

The $2\nu\beta\beta$-decay half-life,
the single electron and 
angular distribution differential decay rates 
within the SSD1, SSD2 and HSD approaches are given as follows:
\begin{eqnarray}
[T^{2\nu -I}_{1/2} (0^{+}\to J^\pi)]^{-1} = \frac{\omega^I}{ln(2)} = 
\frac{c_{2\nu}}{ln(2)}
\int\limits_{m_e}^{E_i - E_f - m_e} k_1 \varepsilon_1  F(Z_f,\varepsilon_1)
d\varepsilon_1 
\times ~~~~~\nonumber \\
\int\limits_{m_e}^{E_i - E_f - \varepsilon_1} k_2 \varepsilon_2 
F(Z_f,\varepsilon_2) d\varepsilon_2
\int\limits_{0}^{E_i - E_f - \varepsilon_1-\varepsilon_2} \nu_1^2 \nu_2^2 
{\cal A}^I_{J^\pi} d\nu_1,~~
\label{eq:7}
\end{eqnarray}

\begin{eqnarray}
\frac{d \omega^I (0^{+}\to J^{\pi})}{d \varepsilon_1} = 
c_{2\nu}
k_1 \varepsilon_1  F(Z_f,\varepsilon_1) \times \nonumber \\
\int\limits_{m_e}^{E_i - E_f - \varepsilon_1} k_2 \varepsilon_2 
F(Z_f,\varepsilon_2) d\varepsilon_2 
&&\int\limits_{0}^{E_i - E_f - \varepsilon_1-\varepsilon_2} \nu_1^2 \nu_2^2 
{\cal A}^I_{J^\pi} d\nu_1,
\label{eq:8}
\end{eqnarray}

\begin{eqnarray}
\frac{d \omega^I (0^{+}\to J^{\pi})}{d \cos{\theta}} &=& 
\frac{c_{2\nu}}{2}
\int\limits_{m_e}^{E_i - E_f - m_e} k_1 \varepsilon_1  F(Z_f,\varepsilon_1)
d\varepsilon_1
\int\limits_{m_e}^{E_i - E_f - \varepsilon_1} k_2 \varepsilon_2 
F(Z_f,\varepsilon_2) d\varepsilon_2
\times \nonumber \\
&&\int\limits_{0}^{E_i - E_f - \varepsilon_1-\varepsilon_2} \nu_1^2 \nu_2^2 
\left({\cal A}^I_{J^\pi} + {\cal B}^I_{J^\pi} 
\frac{k_1 k_2}{\varepsilon_1 \varepsilon_2}
\cos{\theta}\right)
d\nu_1.
\label{eq:9}
\end{eqnarray}
Here, $c_{2\nu} = G^4_\beta g_A^4 / 8 \pi^7$ and
$F(Z_f,\varepsilon )$ is the relativistic Coulomb factor \cite{HAX84,DOI85}.
The expressions for the 
factors ${\cal A}^I_{J^\pi}$ and ${\cal B}^I_{J^\pi}$ ($I = SSD1$, 
$SSD2$, $HSD$ 
and $J^\pi = 0^+, ~2^+$) are presented in Table \ref{table.1}.

In order to determine 
the $2\nu\beta\beta$-decay half-life 
within the SSD hypothesis the matrix elements $M^f_1(J^+)$ 
and $M^i_1(0^+)$ have to be
specified. They can be deduced from the $log ft$ values of  electron capture 
and the single $\beta$ decays as follows:
\begin{equation}
M^i_1(0^+) = \frac{1}{g_A}\sqrt{\frac{3 D}{ ft_{EC}}}, ~~~
M^f_1(J^+) = \frac{1}{g_A}\sqrt{\frac{3 D}{ ft_{\beta^-}}}.
\label{eq:10}
\end{equation}
Here, $D= (2 \pi^3 \ln 2)/(G^2_\beta m_e^5)$ 
($G_\beta = 1.149\times 10^{-5} GeV^{-2}$)
and $g_A$ is the 
vector-axial coupling constant. The advantage of this phenomenological
determination of the beta transition amplitudes $M^i_1(0^+)$ and $M^f_1(J^+)$ 
consists in their nuclear model independence and in the fact that the 
associated $2\nu\beta\beta$-decay rate does not depend explicitly on $g_A$
($g_A$ factors from beta amplitudes in Eq. (\ref{eq:10}) are canceled
with  $g^4_A$  from $c_{2\nu}$ factor).

\section{Calculation and Discussion}
\label{sec:level3}

In this paper we study the SSD hypothesis for $2\nu\beta\beta$-decay
of $^{100}Mo$ to the ground state as well as to the $0^+$ and $2^+$
excited states of the final nucleus $^{100}Ru$.  
The calculated $2\nu\beta\beta$-decay half-lifes are presented in 
Table \ref{table.2} and compared with the available experimental data. 
By comparing the results of the SSD1 and SSD2 approaches we see that the
exact consideration of the energy denominators  leads to a significant
reduction of the half-lifes for all studied transitions and that 
this effect is especially large for transitions
to the $2^+$ states of the final nucleus. 
The obtained SSD2 values are close to the experimental  
ones both  for the transition to the ground  and
excited $0^+_1$  states of 100Ru. However, it is not possible 
to draw a general conclusion with respect to the SSD approach as there
is some disagreement between different experimental 
measurements (see Table \ref{table.2}). 
We note also that the phenomenological predictions
for the 2vbb-decay half-lives (SSD1 and SSD2) have a big uncertainty too
($\approx 50$ \%) due to inaccurate experimental determination of the 
$log ft_{EC}$ value for the electron capture \cite{bar99}.  
It is expected that the above drawbacks will be eliminated by
the future experimental measurements \cite{bar99}.

Till now the $2\nu\beta\beta$-decay transition to the $2^+$ state
of the final nucleus has been not observed. The SSD2 results given
in Table \ref{table.2} suggest that this transition for the A=100 system
can be detected on the level of $10^{23}$ years. In spite of the 
advantage of $2\nu\beta\beta$-decay measurement to $2^+$ excited state 
in coincidence with gamma transition to
the $0^+$ ground state the detection of this $2\nu\beta\beta$-decay transition
seems to be unreachable in near future. We note that our calculation 
has been performed by considering that all outgoing leptons are
emitted in the S-wave state. This case is favored from the viewpoint
of lepton wave but suppressed due to the small factor $({\cal K}-{\cal L})$.
However, there are other possibilities from the higher partial
waves of leptons with favored additive combination $({\cal K}+{\cal L})$
of denominators. In Ref. \cite{tom91} it was estimated that the suppression
factor due to a P-wave to S-wave ratio is about $10^{-3}$ for the
of $2\nu\beta\beta$-decay amplitude. We have found that the contribution 
from higher lepton partial wave can be comparable  with the pure S-wave 
contribution only in the case in which the above ratio is about 50,  something
unexpected. Thus within the SSD hypothesis the 
$2\nu\beta\beta$-decay transition to the $2^+$ state is governed by
the pure lepton S-wave contribution. 

Further, we have found that it is possible experimentally decide whether 
one low-lying state dominates or not by precise measuring the single electron 
spectra and/or angular distributions. The single electron spectrum of the
emitted electrons calculated within SSD (i.e.,  SSD2) and HSD approaches
is shown in Fig. \ref{fig.1}. The SSD and HSD distributions associated
with the transitions to the $0^+$ ground (Fig. \ref{fig.1}a)
and excited  (Fig. \ref{fig.1}b) states of the final
nucleus were normalized to the experimental half-lifes of Ref.
\cite{sil97} and Ref. \cite{barb95} (see Table \ref{table.2}), 
respectively. As there are no available $2\nu\beta\beta$-decay data
for the transition to the $2^+_1$ excited state of the final nucleus,
the distributions in Fig. \ref{fig.1}c were normalized to the 
half-life predicted by the SSD2 approach (see Table \ref{table.2}).
By glancing Figs. \ref{fig.1} we see that there is different behavior
of the single electron differential decay rate calculated within
SSD (i.e., SSD2) and HSD approaches especially for small electron energy.
It is supposed that this SSD versus HSD effect is enough large to be studied  
by the NEMO III experiment, which is currently in preparation \cite{bar99}. 

The NEMO III experiment is supposed to achieve precise measurement 
of the angular correlation of outgoing electrons as well.  
The curves representing the SSD (i.e., SSD2) and HSD approaches 
for this observable characteristic are just lines with 
different asymptotic behavior [see Eq. \ref{eq:9}]. We have 
\begin{equation}
\frac{d \omega^I (0^{+}\to J^{\pi})}{d \cos{\theta}} =
\frac{1}{2} \omega^I (0^{+}\to J^{\pi})~ [1~ + ~\kappa^I (0^{+}\to J^{\pi}) 
~\cos{\theta} ~].
\end{equation}
In the case of $2\nu\beta\beta$-decay of ${100}Mo$ one obtains
\begin{eqnarray}
\kappa^I (0^+\to 0^+_{g.s.}) &=& -0.627  ~(I=SSD),  
~~~  -0.646 ~(I=HSD) \nonumber \\
\kappa^I (0^+\to 0^+_{1}) &=&  -0.487 ~(I=SSD),  
~~~  -0.450 ~(I=HSD) \nonumber \\
\kappa^I (0^+\to 2^+_{1}) &=&  ~0.153  ~(I=SSD),  
~~~  ~0.149~(I=HSD) 
\end{eqnarray}
We see that there is only a small difference 
between the SSD and HSD values of $\kappa^I$. 
Nevertheless, it is expected that this effect can be tested by the 
NEMO III experiment too \cite{bar99}. 

We maintain that the study of $2\nu\beta\beta$-decay differential 
characteristics offers a new possibility to 
decide whether one low-lying state dominates or not. It
is more reliable way as  a simple comparison of calculated and
measured half--lifes.

\section{Summary and Conclusions}
\label{sec:level4}

We have studied the $2\nu\beta\beta$-decay of $^{100}Mo$
in the context of the SSD hypothesis. To our knowledge,
the validity of the separation of the lepton and nuclear parts 
has been discussed for the first time. The 
transitions to the ground  ($0^+$) and excited 
($0^+$ and $2^+$) states of the final nucleus has been considered.
We have shown
that by exact treatment of the lowest state of the intermediate
nucleus the $2\nu\beta\beta$-decay half--life is reduced
by factor of $20$ percent for transitions to $0^+$ states.
However, much larger reduction appears for $2\nu\beta\beta$-decay 
transitions to $2^+$ states amounting to $300$ percent
(see Table \ref{table.2}). In addition,
we have found that the emitted
electrons in these $2\nu\beta\beta$-decay transitions are
predominantly in the S-wave states in the case the
SSD dominance is realized.

\begin{figure}
\centerline{\epsfig{file=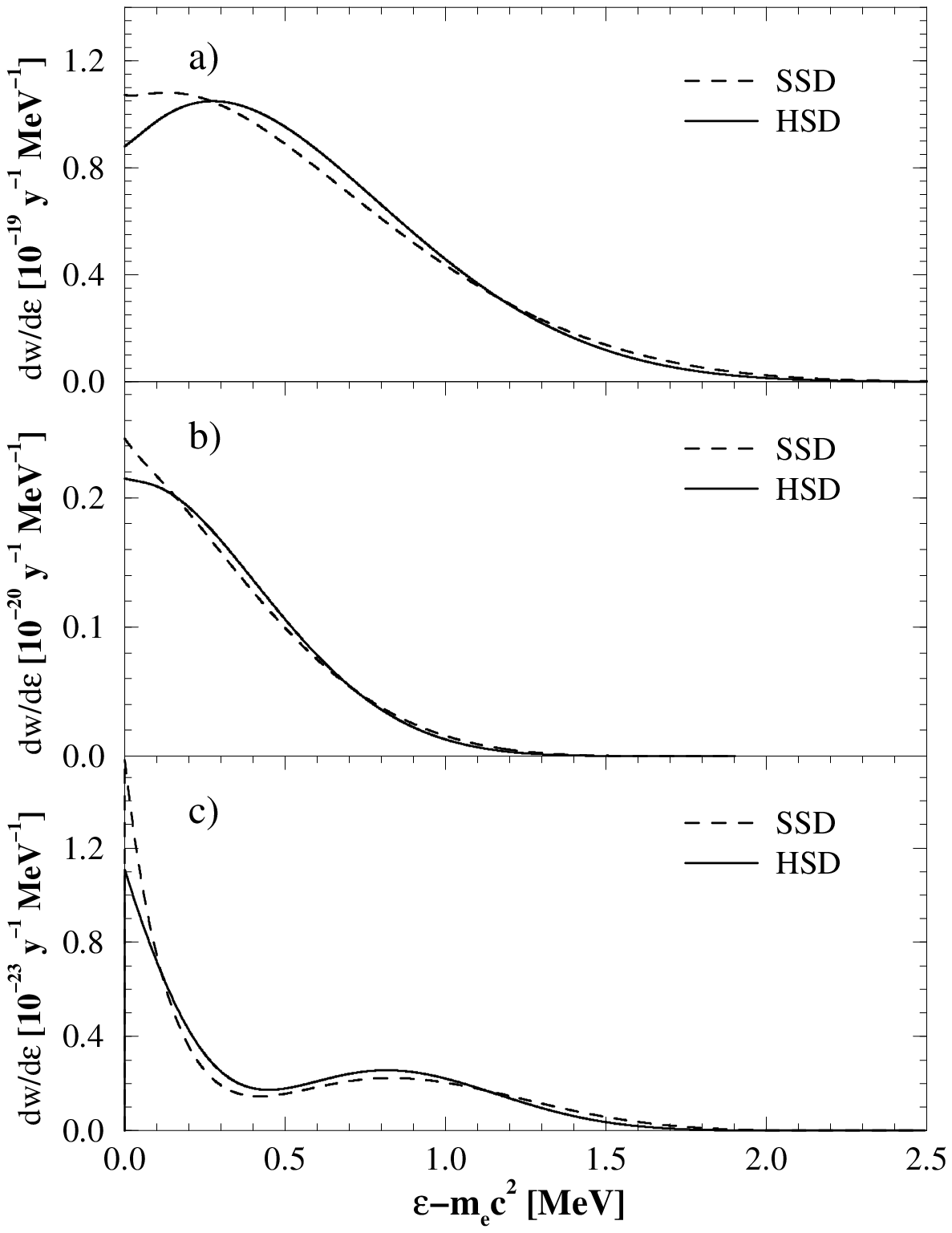,width=14.cm}}
\caption{Single electron differential decay rate ${d w}/{d \epsilon}$
for the $2\nu\beta\beta$-decay  of $^{100}Mo$ to the $0^+$ ground (a),
the first $0^+_1$ excited   (b) and the first $2^+$ excited (c) states 
in $^{100}Ru$. $\epsilon$ and $m_e$ represent the energy and mass 
of the electron, respectively. The calculations have been performed within
the single-state dominance hypothesis 
(SSD2-exact calculation) and by assuming the dominance of higher lying
states (HSD).
}
\label{fig.1}
\end{figure}

Further, we have shown that one can learn more
about details of the $2\nu\beta\beta$-decay nuclear transition
by measuring the single electron spectra and/or angular distributions
of the emitted electrons. We have found that
the SSD and the HSD differential decay--rates  
exhibit different behaviour (see Fig. \ref{fig.1}). 
It is expected that this SSD versus HSD effect can be studied experimentally, 
e.g. by the NEMO III collaboration \cite{bar99}, 
which has the chance to confirm or rule out the SSD
hypothesis in the near future. This kind of information is expected
to be very helpful in understanding the details of nuclear structure.

{\it Acknowledgments}. 
The authors are grateful to A.S. Barabash, Yu.V. Gaponov and G. Pantis
for useful discussions. We acknowledge a support by Slovak  
VEGA grant $V2F31_G$ and
the Grant Agency of the Czech  Republic grant No. 202/98/1216.


\newpage

\begin{table}[t]
\caption{The functions ${\cal A}^I_{J^\pi}$ and ${\cal B}^I_{J^\pi}$ entering 
the expressions for the $2\nu\beta\beta$-decay rate in Eqs. (7), (8) 
and (9) within the SSD1, SSD2 and HSD hypothesis. Here, 
${\cal K} = {1}/{D(\varepsilon_1,\omega_1)} +
{1}/{D(\varepsilon_2,\omega_2)}$ and  
${\cal L} = {1}/{D(\varepsilon_1,\omega_2)} +
{1}/{D(\varepsilon_2,\omega_1)}$ with
$D(\varepsilon_i,\omega_j)\equiv {E_1 - E_i + \varepsilon_i +\nu_j }$ 
(i,j = 1,2).
} 
\begin{tabular}{lccc}
I & SSD1 & SSD2 & HSD \\ \hline
${\cal A}^I_{0^+}$ &
$\frac{|M^f_1(0^+) M^i_1(0^+)|^2}{(E_1 - E_i + \Delta )^2}$ &
$| {M^f_1(0^+) M^i_1(0^+)} |^2~~
\frac{{\cal K}^2 + {\cal L}^2 + {\cal K}{\cal L}}{12}$ &
$| M^{exp}_{GT} (0^+) |^2 $ \\
${\cal B}^I_{0^+}$ &
$- \frac{|M^f_1(0^+) M^i_1(0^+)|^2}{(E_1 - E_i + \Delta )^2}$ &
$- | {M^f_1(0^+) M^i_1(0^+)} |^2~~
\frac{2{\cal K}^2 + 2{\cal L}^2 + 5{\cal K}{\cal L}}{36}$ &
$- | M^{exp}_{GT} (0^+) |^2 $ \\
${\cal A}^I_{2^+}$ &
$\frac{|M^f_1(2^+) M^i_1(0^+)|^2}{3(E_1 - E_i + \Delta )^6}\times $ &
$ \frac{|{M^f_1(2^+) M^i_1(0^+)}|^2}{3}~~ 
\frac{({\cal K}-{\cal L})^2}{4}$ &
$| M^{exp}_{GT} (2^+) |^2 \times$ \\ 
 & $(\varepsilon_1-\varepsilon_2)^2 (\nu_1-\nu_2)^2$ & &
$(\varepsilon_1-\varepsilon_2)^2 (\nu_1-\nu_2)^2 $\\ 
${\cal B}^I_{2^+}$ &
$\frac{|M^f_1(2^+) M^i_1(0^+)|^2}{3(E_1 - E_i + \Delta )^6} \times$ & 
$ \frac{|{M^f_1(2^+) M^i_1(0^+)}|^2}{3}~~ 
\frac{({\cal K}-{\cal L})^2}{12}$ &
$| M^{exp}_{GT} (2^+) |^2 \times$ \\ 
 & $\frac{(\varepsilon_1-\varepsilon_2)^2 (\nu_1-\nu_2)^2}{3}$ & &
$\frac{(\varepsilon_1-\varepsilon_2)^2 (\nu_1-\nu_2)^2}{3} $
\end{tabular}
\label{table.1}
\end{table}

\begin{table}[t]
\caption{Calculated half-lifes for the $2\nu\beta\beta$-decay transitions
from  the ground state  of $^{100}Mo$ to the ground state ($0^+_{g.s.}$)  
and excited states ($0^+_1$ and  $2^+_k$, k=1,2)
of $^{100}Ru$ within the single-state 
dominance hypothesis with approximated (SSD1) and exact (SSD2) $\cal{K}$ and 
$\cal{L}$ factors. $T_{1/2}^{2\nu -exp}$ is the experimental half-life and
$W_{if}=E_i-E_f$ is the energy difference of initial and final nuclei.
We considered $\log ft_{EC}$ to be 4.45 
\protect\cite{gar93}. The product of matrix elements
$M^I_{1}M^F_{1}$ is calculated for $g_A=1.25$.
} 
\begin{tabular}{lcccccc}
Transition  & $W_{if}$   & $\log ft_{\beta^-}$ & 
$M^I_{1}M^F_{1}$  & $T_{1/2}^{2\nu -SSD1}$  &$T_{1/2}^{2\nu -SSD2}$  & 
$T_{1/2}^{2\nu -exp}$  [Ref.]\\
 & (MeV) &  &  & (y) & (y) & (y) \\
\hline
$0^{+}_{g.s.} \to 0^{+}_{g.s.}$
& $4.057$ & $4.6$ & $0.352$ & 
$8.97\times 10^{18}$ & $7.15\times 10^{18}$ & 
$(6.82^{+0.38}_{-0.53}\pm 0.68)\times 10^{18}$ \cite{sil97}\\
& & & & & & $(9.5\pm 0.4\pm 0.9)\times 10^{18}$ \cite{das95}\\
& & & & & & $(11.5^{+3.0}_{-2.0})\times 10^{18}$ \cite{eji91}\\
& & & & & & $(7.6^{+2.2}_{-1.4})\times 10^{18}$ \cite{als97}\\
$0^{+}_{g.s.} \to 0^{+}_{1}$
& $2.926$ & $5.0$ & $0.222$ & 
$5.44\times 10^{20}$&$4.45\times 10^{20}$  &
$(6.1^{+1.8}_{-1.1})\times 10^{20}$ \cite{barb95}\\
$0^{+}_{g.s.} \to 2^{+}_{1}$
& $3.517$ & $6.5$ & $0.0395$ & 
$ 4.66\times 10^{23}$ & $1.73\times 10^{23}$ & $>16\times 
10^{20}$ \cite{barb95}\\
$0^{+}_{g.s.} \to 2^{+}_{2}$ & 
$2.694$ & $7.1$ & $0.0198$ & $3.34\times 10^{25}$ & 
$1.45\times 10^{25}$ & $> 13\times 10^{20}$ \cite{barb95}
\end{tabular}
\label{table.2}
\end{table}

\end{document}